\documentclass{aa}  

\usepackage{graphicx}
\usepackage{txfonts}

\begin{document}

   \title{Modelling the 2020 November 29 solar energetic particle event using the EUHFORIA and the iPATH model }


   \author{Zheyi Ding
          \inst{1},
          Nicolas Wijsen\inst{1,2,3},
           Gang Li\inst{4}, \and  Stefaan Poedts \inst{1,5}
          }

   \institute{\inst{1} Centre for mathematical Plasma Astrophysics, KU Leuven, 3001 Leuven, Belgium; \email{stefaan.poedts@kuleuven.be}\\
              \inst{2} NASA, Goddard Space Flight Center, Heliophysics Science Division, Greenbelt, MD 20771, USA \\
              \inst{3} Department of Astronomy, University of Maryland, College Park, MD 20742, USA \\
              \inst{4} Department of Space Science and CSPAR, University of Alabama in Huntsville, Huntsville, AL 35899,USA;
             \email{gangli.uahuntsville@gmail.com}\\
             \inst{5} Institute of Physics, University of Maria Curie-Sk{\l}odowska, Pl.\ M.\ Curie-Sk{\l}odowska 5, 20-031 Lublin, Poland
             }


 
  \abstract
   {}
   {We present the implementation of coupling the EUropean Heliospheric FORcasting Information Asset (EUHFORIA) and the improved Particle Acceleration and Transport in the Heliosphere (iPATH) model and simulate the widespread solar energetic particle (SEP) event of 2020 November 29. We compare the simulated time intensity profiles with measurements at Parker Solar Probe (PSP), the Solar Terrestrial Relations Observatory (STEREO)-A, SOlar and Heliospheric Observatory (SOHO) and Solar Orbiter (SolO). We focus on the influence of the history of shock acceleration on the varying SEP time intensity profiles and investigate the underlying causes in the origin of this widespread SEP event.    }
   {We simulate a magnetized coronal mass ejection (CME) propagating in the data-driven solar wind with the EUHFORIA code. The CME is initiated by using the linear force-free spheromak module of EUHFORIA. Shock parameters  and a 3D shell structure are computed from EUHFORIA as inputs for the iPATH model. Within the iPATH model, the steady-state solution of particle distribution assuming diffuse shock acceleration is obtained at the shock front. The subsequent SEP transport is described by the focused transport equation using the backward stochastic differential equation method with perpendicular diffusion included.}
   {The temporal evolution of shock parameters and particle  fluxes during this event are examined. We find that adopting a realistic solar wind background can significantly impact the expansion of the shock and consequently the shock parameters. Time intensity profiles with an energetic storm particle event at PSP are well reproduced from the simulation. In addition, the simulated and observed time intensity profiles of protons show a similar two-phase enhancement at STA. These results illustrate that modelling a shock using a realistic solar wind is crucial in determining the characteristics of SEP events. The decay phase of the modelled time intensity profiles at Earth agrees well with observations, indicating the importance of perpendicular diffusion in widespread SEP events. Taking into account the possible large curved magnetic field line connecting to SolO, the modelled time intensity profiles show good agreement with the observation. We suggest that the largely distorted magnetic field lines due to a stream interaction region may be a key factor in understanding the observed SEPs at SolO in this event.}
   {}

   \titlerunning {Modelling the 2020 November 29 SEP event}
   \authorrunning{Ding et al.}
   \maketitle

%

\section{introduction}
Large solar energetic particle (SEP) events have been a major concern in space weather since the accompanying high energy ($>10$ MeV) protons and ions pose serious radiation threats to astronauts and satellites in space \citep{Desai2016}. These SEP events are typically associated with shock waves driven by coronal mass ejections (CMEs), and are classified as gradual events \citep{Cane+1988,Reames+1999}. Comparing to the other category of SEP events, which are associated with solar flares and are known as impulsive events,  gradual SEP events exhibit larger spatial spreading of SEPs and over longer timescales. This is because the CME-driven shock can accelerate particles over a large angular extent across the shock front. Multiple spacecraft observations from well-separated longitudes have revealed that the angular spread of SEP can be even more than $180^{\circ}$ \citep{Gomez2015ApJ...799...55G,Lario2017ApJ...847..103L}. Several processes have been proposed to explain these observations. Coronal shocks can accelerate and inject particles into a large spatial extension near the sun \citep{Dresing2014}. Cross-field transport in the corona and the interplanetary medium can also lead to widespread SEP events \citep[e.g.,][]{Hu+etal+2017,Li2021,Wijsen2019JPhCS1332a2018W}. Other mechanisms, including a diverging coronal magnetic field \citep{Klein2008A&A...486..589K} and the expansion of EUV waves \citep{Park2013ApJ...779..184P},  have also been proposed to explain the spreading of SEPs.

 Understanding particle acceleration at the shock and subsequent  transport is essential for investigating widespread SEP events. Diffusive shock acceleration (DSA), also known as first-order Fermi acceleration, is believed to be the primary particle acceleration process at CME-driven shocks. In the DSA mechanism, particles are accelerated by continuously traversing the shock via scattering between Alfv\'{e}n waves upstream and downstream of the shock.  Accelerated particles streaming along the magnetic field line upstream of the shock further amplify these Alfv\'{e}n waves. To apply the DSA in interplanetary shocks, \citet{Lee+1983} and \citet{Gordon+1999} obtained the spatial diffusion coefficient by solving the coupled particle diffusion equations and upstream Alfv\'{e}n wave kinetic equations. These solutions were adopted by \citet{Zank+etal+2000, Rice+2003, Li+2003} in modelling gradual SEP events. The original model developed by \citet{Zank+etal+2000} used the one-dimensional (1D) ZEUS Magnetohydrodynamics (MHD) code \citep{Clarke1996} to model the solar wind and a CME-driven shock. The time-dependent maximum cut-off energy at the shock front is determined by equating the dynamical timescale of the shock and the acceleration time scale \citep{Drury+1983}. The instantaneous steady-state particle spectrum with a cut-off energy is adopted in the model. In this work, \citet{Zank+etal+2000} developed a treatment, referred to as the onion shell model, to track the convection and diffusion of accelerated particles in the shock complex. Later \citet{Rice+2003} extended the work of  \citet{Zank+etal+2000} by including the consideration of shocks with arbitrary strength and \citet{Li+2003} further improved the original model using a Monte-Carlo approach to model the transport of escaped particles and \citet{Li+2005a} included the treatment of heavy ions in the model. These works converged into a comprehensive model,  named the Particle Acceleration and Transport in the Heliosphere (PATH) model, which was a 1D model. Shortly after,  using the PATH model, \citet{Verkhoglyadova+2009,Verkhoglyadova+2010} successfully modelled the time intensity profiles and spectra of several SEP events. Due to the fact it was a 1D model, the PATH model can not describe particle acceleration at a quasi-perpendicular or an oblique shock. As a workaround, \citet{Li+2012} examined the effect of shock obliquity using the PATH model by considering a shock with a fixed obliquity during its propagation. This is of course, not realistic since the shock obliquity can vary significantly during its propagation. A self-consistent two-dimensional extension of the PATH model was done by \citet{Hu+etal+2017}, resulting  the improved Particle Acceleration and Transport in the Heliosphere (iPATH) model, which follows the shock obliquity self-consistently. Using the iPATH model, \citet{Hu+etal+2018} examined time intensity profiles and particle spectra at different longitudes and radial distances for some example SEP events.  More recently, \citet{Ding+2020} has examined the 2017 September 10 Ground Level Enhancement (GLE) event using the iPATH model. 
In a different context, \citet{Fu+2019} modelled stellar energetic particle events with different stellar rotation rates using iPATH, and \citet{hu2022extreme} derived the scaling relation between energetic particle fluence and CME speed in stellar energetic particle events. 

While applying PATH and iPATH to model individual SEP events has been done successfully in many cases, one unsatisfying issue with these practices is the fact that the iPATH model uses a uniform solar wind background and adopts a simplified approach in initiating a CME-driven shock. This is inconsistent with realistic SEP events where the background solar wind is non-uniform, often affected by transient structures, including CMEs erupted earlier and stream interaction regions (SIRs).  Such a non-uniform background solar wind can impact the acceleration and the transport process of SEPs and lead to particle spectra and time intensities very different from the cases with a uniform background.  Therefore, it is essential to couple data-driven solar wind models and more sophisticated CME models in SEP simulations.

Previous efforts that couple data-driven three-dimensional (3D) MHD simulations of the solar wind and CMEs with SEP models exist. For example, \citet{Kozarev2013}  coupled the Block Adaptive Tree Solar-Wind Roe Upwind Scheme model (BATSRUS; \citep{toth2012JCoPh.231..870T}) with the Earth-Moon-Mars Radiation Environment Module (EMMREM; \citet{Schwadron+2010}) to simulate particle acceleration at CME-driven shock in the solar corona. Similarly,  \citet{Linker+2019} and \citet{Young2021} have coupled the CORHEL (Corona-Heliosphere; \citet{Riley+2012}) and the EMMREM to investigate the July 14 2000 SEP event. \citet{Luhmann+2007,Luhmann+2010} developed a SEP model, SEPMOD, which uses the outputs of the WSA-ENLIL-cone model in describing the interplanetary field lines. The connection of an observer as a function of time and the shock properties along the field line is obtained from the WSA-ENLIL model. This approach emphasizes the importance of the magnetic field connection between the observer and the shock. Recently, \citet{wijsen2019modelling} developed a three-dimensional particle transport model, the PArticle Radiation Asset Directed at Interplanetary Space Exploration (PARADISE) model, that solves the focused transport equation by coupling to the EUropean Heliospheric FORecasting Information Asset (EUHFORIA; \citet{pomoell2018euhforia}). EUHFORIA is a data-driven MHD model that simulates solar wind and CME propagation in the inner heliosphere. The EUHFORIA+PARADISE model successfully reproduced an energetic particle event at the corotating interaction region (CIR) and an energetic storm particle (ESP) event at interplanetary CME-driven shock \citep{wijsen2021observation,wijsen2021self}. However, 
these models did not consider the time-dependent wave amplification upstream of the shock, which is essential in obtaining the maximum energy of the accelerated particles.  \citet{Li2021} combined the iPATH model and the Alfv\'en Wave Solar Model (AWSoM; \citet{vanderholst10}) to simulate the 2012 May 17 GLE event. They investigated the particle acceleration at a 3D CME-driven shock below 30 solar radii and obtained good agreement between simulations and observations at multiple spacecraft. This modeling effort showed the power of the iPATH model when coupled with data-driven MHD models in studying the characteristics of particle acceleration at CME-driven shocks.

In this work, we combine the EUHFORIA code with the iPATH model to investigate the 2020 November 29 SEP event. This event is the first widespread SEP event of solar cycle 25 observed by Solar Orbiter (SolO), Parker Solar Probe (PSP), Solar Terrestrial Relations Observatory-A (STA), and multiple missions near the Earth. We first use the EUHFORIA code to model the background solar wind and the propagation of the CME from 0.1 au to 2 au. Remote sensing observations of the CME and in-situ plasma measurements of the shock arrivals at STA and PSP are used to constrain the spheromak CME parameters in EUHFORIA. The shock parameters from the EUHFORIA are taken as the input of the iPATH model. The iPATH model solves the particle distribution function at the shock front and the escaped particles from the shock front are tracked in the transport module. The modelled time intensity profiles of $>10$ MeV protons at four spacecraft are obtained and compared with the observations. 

The 2020 November 29 SEP event has attracted much attention in the SEP community, and was discussed in an overview of PSP SEP observations by \citet{cohen2021psp}. 
This event originated from a fast and relatively wide CME associated with an M4.4 class X-ray flare from the active region (AR) 12790. The event showed an unusual wide spreading with particles observed  at locations with at least 230$^{\circ}$ difference in longitude. Large anisotropies are observed by SolO, STA and PSP at the onset of this event \citep{cohen2021psp,kollhoff2021first}, indicating that the injection of particles near the Sun possibly extended over a wide longitudinal range.  \citet{kouloumvakos2022first} examined the magnetic field configurations in the low corona and showed that three spacecraft (SolO, STA and PSP) connect to supercritical regions of the shock. They suggested that shock properties play an important role in this widespread SEP event. In addition, \citet{kollhoff2021first} suggested that efficient cross-field transport might have occurred in this event. Besides the wide spreading of SEPs, other interesting characteristics of this SEP event have been discussed, such as the significant depletion of SEPs associated with the interaction between the shock and a magnetic structure by \citet{giacalone2021energetic}, the similarity and differences of ion spectra at four spacecraft by \citet{mason2021solar}, and the observation of inverted energy spectra due to the passage of previous CME  \citep{lario2021comparative}. More recently, the observation of this event near Mars as observed by Tianwen-1 has been reported by \citet{fu2022ApJ...934L..15F}. Despite the fact that many observational studies have been devoted to this event, so far comprehensive modelling effort on this SEP event is still lacking. Very recently, \citet{palmerio2022cmes} simulated the CME and the SEP event using the WSA–Enlil–SEPMOD modelling combo. Their model calculation suggested that significant enhancements of SEP intensity at the onset of the event only exist at PSP. This, however, is because the SEPMOD model assumes scatter-free propagation along magnetic field lines and no cross-field diffusion is included. If, on the other hand, the cross-field diffusion is important, and it is being the key factor leading to the wide-spreading nature of this event,  then including it into the modeling effort is essential. In this work, we focus on modelling the time intensity profiles at four spacecraft and understanding the nature of the widespread SEPs in the 2020 November 29 SEP event. 

Our paper is organized as follows. In Sect.~\ref{sec:model}, we introduce the coupling between the EUHFORIA and the iPATH model, including various parameters of the spheromak module in EUHFORA, the extraction of the shock parameter from EUHFORA and passage of them to iPATH. We also briefly describe the treatment of shock acceleration and particle transport in the iPATH model. 
Section~\ref{sec:results}  contains the MHD simulation results, the history of shock acceleration and the comparisons of time intensity profiles between our simulations and observations. We also compare the simulated results  with and without cross-field diffusion. The main conclusions of this work are summarized in Sect.~\ref{sec:conclu}.

\section{Model Setup}\label{sec:model}
\subsection{Modelling the solar wind and CME with EUHFORIA}

EUHFORIA is a data-driven coronal and heliospheric model designed for space weather forecasting, which simulates the realistic background solar wind and CMEs in the inner heliosphere. This model consists of two major modules: (1) the coronal model, which utilizes synoptic magnetograms from the Global Oscillation Network Group (GONG; \citet{harvey1996global}) as inputs for the semi-empirical Wang-Sheeley-Arge-like model (WSA; \citet{arge2003improved}). The empirical coronal model can provide inner boundary conditions for MHD models at 0.1 au. (2) the heliospheric model, which solves the 3D time-dependent MHD equations to generate a realistic background solar wind and CMEs. Several CME models have been implemented in EUHFORIA, including the cone model \citep{zhao2002determination,odstrcil2004numerical} and the linear force-free spheromak (LFFS) model \citep{verbeke2019evolution}. In the cone model, CMEs are described as impulsive hydrodynamic disturbances of plasma propagating into the solar wind with a constant insertion speed and angular width. When the cone CMEs are injected into the heliosphere, where their propagation and deformation are affected by the internal pressure of the injected solar wind gust and the ambient solar wind structures. However, the cone model can not trace the additional expansion of the magnetic structure in the CME. Compared to the cone model,  the LFFS model contains a flux-rope structure as a force-free magnetic field configuration, which can more realistically describe the propagation and evolution of CMEs in the solar wind \citep{scolini2019observation}. 

In this work, we first model the background solar wind with a spheromak CME in EUHFORIA. The remote-sensing and in-situ plasma measurements are used to constrain the parameters of the LFFS model. Using the coronagraphs onboard STA and SOlar and Heliospheric Observatory (SOHO), \citet{nieves2022direct} obtained kinematic CME parameters at 0.1 au from the Graduated Cylindrical Shell (GCS; \citet{thernisien2009forward}) reconstructions for the ENLIL model. We follow these GCS parameters in this work. Besides the remote-sensing, in-situ magnetic field measurements provide information on the arrival of the shock and CME. The CME-driven interplanetary shock reached PSP (at 0.81 au from the Sun) at 18:35 UT on 30 November 2020 and reached STA at 07:23 UT on 1 December 2020. By fine-tuning the magnetic flux and density to match the shock arrival time at two SC locations, the reasonable CME parameters for the LFFS model are summarised in Table~\ref{table1-lffs}. Grid resolutions in EUHFORIA are as follows: 1024 grid cells in the radial direction between 0.1 au and 2.0 au, and a $4^{\circ}$ angular resolution in longitudes and latitudes.

\begin{table}
\caption{\label{table1-lffs}Input parameters of the spheromak CME model in the EUHFORIA}
\centering
\begin{tabular}{lccc}
\hline\hline
Parameter &Value\\
\hline
Insertion time           &  2020-11-29T15:15:00 \\
Insertion longitude (HEEQ)      & -15$^\circ$  \\
Insertion longitude (HEEQ) &  -80$^\circ$ \\
Radius          &  21.5 $R_{\odot}$\\
Density     &   $3.0$ $\times$ $10^{-19}$ kg m$^{-3}$\\
Temperature      &  $0.8$ $\times$ $10^6$ \\
Helicity      &  $1.0$  \\
Tilt      &   $-70^\circ$ \\
Toroidal magnetic flux      &  $1.5$ $\times$ $10^{14}$ Wb \\
\hline
\end{tabular}
\end{table}

\subsection{Shock parameters}
In order to study CME-driven shocks in EUHFORIA simulations, the shock structure has to be identified exactly. To diminish the effect of the non-uniform solar wind on the shock identification, at every time step, we first subtract the background solar wind that corotates with the Sun, which enables us to remove any disturbance in front of CME. Next, we identify the shock front along the radial direction with the last peak of entropy $s$ at a series of times. The entropy is calculated by $s=ln(T_{p}/\rho^{\gamma-1})$ where $T_{p}$ is the proton temperature and $\gamma = 5/3$ is the polytropic index. With a 2-hour cadence for the snapshots in EUHFORIA, we can track the shock parameters (e.g., the compression ratio $S$, shock speed $V_{\rm shock}$, and shock obliquity angle $\theta_{\rm BN}$) during the CME evolution. The shock obliquity angle $\theta_{\rm BN}$ refers to the angle between the shock normal and the upstream magnetic field. At each shock location, the shock normal is determined by using magnetic coplanarity \citep{abraham1972determination}:
   \begin{equation}
\mathbf{n}=\pm \frac{\left ( \mathbf{B}_{d} \times \mathbf{B}_{u}\right )\times \left ( \mathbf{B}_{d} -\mathbf{B}_{u} \right ) }{|\left ( \mathbf{B}_{d} \times \mathbf{B}_{u}\right )\times \left ( \mathbf{B}_{d} -\mathbf{B}_{u} \right )|},
   \end{equation}
where $\mathbf{B}_{u}$ and $\mathbf{B}_{d}$ represent upstream and downstream magnetic fields respectively. 

It is supposed that the upstream and downstream speeds are radial, the following relation \citep{whang1996interplanetary} is used in order to estimate shock speed in the shock normal direction:
   \begin{equation}
V_{\text{shock}} = \frac{\rho_{d}u_{d}-\rho_{u}u_{u}}{\rho_{d}-\rho_{u}},
   \end{equation}
where $\rho_{d}$ , $\rho_{u}$ ,$u_{u}$ and $u_{d}$ are the proton density and flow velocity in the upstream and downstream regions, respectively. We then determine the shock parameters (e.g., the shock compression ratio and shock obliquity angle) accordingly.

\subsection{iPATH model}
 The original 2D iPATH model contains three modules: (1) an MHD module simulates the background solar wind and the CME-driven shock; (2) an acceleration module computes particle spectra at the shock front; and (3) a transport module follows the propagation of particles escaping upstream of the CME-driven shock.  
For a detailed discussion of the iPATH model, the readers are referred to \citet{Hu+etal+2017,Ding+2020}.  In \citep{Li2021}, the authors tried to couple the AWSoM MHD code with the iPATH code to provide a more realistic description of the CME-driven shock.  Here in this work, we use EUHFORIA to replace the MHD module of the iPATH code.  As in \cite{Li2021}, the acceleration and transport of SEPs are explicitly 3D in nature.
 The shock parameters calculated from EUHFORIA are passed to the acceleration module. The maximum particle momentum $p_{\rm max,\bf{r}}$ along the shock front at every time step is calculated by equating the dynamical timescale of the shock $t_{\rm dyn}=\frac{R}{dR/dt}$  with the acceleration time scale \citep{Drury+1983},
  \begin{equation}
t_{\rm dyn} = \int_{p_{\rm inj,\bf{r}}}^{p_{\rm max,\bf{r}}}\frac{3s_{\bf{r}}}{s_{\bf{r}}-1}\frac{\kappa_{\bf{r}}}{U_{\bf{r}}^{2}}\frac{1}{p}dp,
\label{eq:dynamic time}
\end{equation}
where $p_{\rm inj,\bf{r}}$ is the injection momentum, $\kappa_{\bf{r}}$ is the particle diffusion coefficient, $U_{\bf{r}}$ is the upstream solar wind speed in the shock frame. The calculations of injection momentum and diffusion coefficient are the same as \cite{Hu+etal+2017}. Since the shock parameters vary with the shock location $\bf{r}$, the maximum particle energy can have a strong variation along the shock front.

 In the original iPATH model \citep{Hu+etal+2017,Hu+etal+2018}, the instantaneous particle distribution function at the shock front is described by a single power law. Recently, \citet{Ding+2020,Li2021} included an exponential tail $\rm exp(-E/E_0)$ at high energy end to account for the finite shock acceleration time and finite shock size. However, the exponential tail can be significantly steep as suggested in \citet{Vainio+2007}. They utilized a more general form of the exponential tail $\rm exp(-(E/E_0)^{\alpha})$ to fit the particle energy spectrum at the shock, where $\alpha$ can be larger than 1. In this work, we consider the instantaneous particle distribution function with such a general exponential tail  $\rm exp(-(E/E_0)^{\alpha})$ as
\begin{equation}
f(\mathbf{r}, p,t_k) = c_1*\epsilon_{\mathbf{r}}n_{\mathbf{r}}p^{-\beta}H[p-p_{\rm inj, \mathbf{r}}] 
\exp\left[-\left(\frac{E}{E_{0,\mathbf{r}}}\right)^{\alpha}\right],
\label{eq:fp}
\end{equation}
where $\beta = 3s_{\bf{r}}/s_{\bf{r}}-1$, $s_{\bf{r}}$ is the shock compression ratio at $\bf{r}$(r,$\theta$,$\phi$), $\epsilon_{\bf{r}}$ is the injection efficiency, $n_{\bf{r}}$ is the upstream solar wind density,
$p_{\rm inj,\bf{r}}$ is the particle injection momentum, $E_{0,\mathbf{r}}$ is the kinetic energy that corresponds to a maximum proton momentum $p_{\rm max,\bf{r}}$ and $\alpha$ is a free parameter to describe the steepness of the exponential tail. In this event, the decay at high energy is rather fast and we use a $\alpha=2$. 
The injection rate is assumed to be $0.5\%$ at the parallel shock. $H$ is the Heaviside function, and $c_1$ is a normalization constant given by
\begin{equation}
c_1 = 1/\int_{p_{\rm inj,\bf{r}}}^{+\infty}p^{-\beta} H[p-p_{\rm inj,\bf{r}}]*\exp\left[-\left(\frac{E}{E_{0,\mathbf{r}}}\right)^{\alpha}\right]d^3p.
\end{equation}

The accelerated particles convect with the shock and diffuse downstream of the shock. In \cite{Li2021}, the shell is composed of multiple historical shock fronts at different time steps that convect with the downstream solar wind. The 3D shell model is divided into multiple small parcels in longitudes and latitudes. In this work, the angular resolution of the shell is 4$^\circ$, consistent with the grid resolution of EUHFORIA. We only consider the evolution of the shell along the radial direction. For the shell located at ($\theta,\phi$), 
the outer edge of the whole shell is the shock front $r_i$ ($i$ is the number of time steps) at time $t_i$,
the radial distance $r_j$ of shell $j$ (j = 1,2,...,i-1) at time $t_i$  is determined by,
\begin{equation}\label{eq:shell_1}
r_{j}(t_{i}) = r_{j}(t_{i-1})+\int_{t_{i-1}}^{t_{i}}u(r_{j}(t_{i-1}+t'),\theta,\phi)dt',
\end{equation}
where $u$ is the solar wind speed at the shell location ($r_j,\theta,\phi$) at successive MHD time steps. 
 In discrete form, Eq.~(\ref{eq:shell_1}) becomes,
\begin{equation}\label{eq:shell2}
r_{j}(t_{i}) = r_{j}(t_{i-1})+ u(r_{j}(t_{i-1}),\theta,\phi)(t_{i}-t_{i-1}).
\end{equation}
Equation~(\ref{eq:shell2}) allows us to construct all parcels in the shell from the outputs of the EUHFORIA model. In this work, we build the 3D shell model using the EUHFORIA outputs with a time interval of two hours. We note that the shell model in iPATH is derived via the realistic 3D shock fronts in EUHFORIA. Equation~(\ref{eq:shell_1}) and (\ref{eq:shell2}) shows that the shape of the shell model is not only determined by the shock fronts but also is affected by the downstream solar wind speed. An example of the 3D shell model is referred to \cite{Li2021}. Particle diffusion and convection in the parcels are followed as in \cite{Hu+etal+2017,Ding+2020}.Particles diffuse far enough upstream of the shock can escape. 
In the iPATH model, the transport of these particles in the solar wind is described  by a focused transport equation. \cite{Hu+etal+2017} solved the focused transport equation using the backward stochastic differential equation (SDE) method with cross-field diffusion included in the Parker spiral magnetic field. Here we follow \cite{Hu+etal+2017} in describing the transport of escaped particles in the solar wind. Previous modelling works (e.g., \citet{Ding+2020,Li2021}) have suggested that cross-field diffusion is a key parameter in explaining wide-spreading SEP events. In the quasi-linear theory (QLT) \citep{Jokipii+1966}, the pitch angle diffusion coefficient $D_{\mu \mu}$ is given by,
\begin{equation}\label{eq:dmiumiu}
D_{\mu \mu } =  \frac{2\pi^{2}\Omega^{2}(1- \mu^2)}{B^{2}v\mu}g^{\rm slab} \left ( k_{\parallel} \right ),
\end{equation}
where $\mu$ is pitch angle cosine, $v$ is particle velocity, $g^{\rm slab} $ is the turbulence power spectrum in the solar wind,  $\Omega = \frac{e B}{\gamma  m}$ is particle's gyrofrequency, and particle's resonant wave number is $k_{\parallel} = \Omega (v|\mu|)^{-1}$. $g^{\rm slab}$ we used here is given by \cite{shalchi+2009+book},
\begin{equation}\label{eq:gslab}
g^{slab}(k_{\parallel})=\frac{C(\nu )}{2\pi} l_{\rm slab}\delta B^{2}_{\rm slab} \left ( 1 + k^{2} l^{2}_{\rm slab} \right )^{-\nu},
\end{equation}
where $l_{\rm slab}$ is the slab bendover scale, set to be $10^{9}$ m. $\delta B^{2}_{\rm slab}$ is the strength of the slab magnetic field and the inertial range spectral index $s=2\nu=5/3$. The normalization factor $C(\nu )$ equals
\begin{equation}\label{eq:cnu}
C(\nu) = \frac{1}{2\sqrt{\pi}}\frac{\Gamma(\nu)}{\Gamma(\nu-1/2)},
\end{equation}
where $\Gamma(x)$ is the Gamma function. A well-known parallel diffusion coefficient $\kappa_{\parallel}$ is,
\begin{equation}\label{eq:kappa_para}
\kappa_{\parallel} = \frac{v^2}{8}\int^{+1}_{-1}d\mu\frac{(1-\mu^2)^2}{D_{\mu\mu}}.
\end{equation}
Then we obtain the perpendicular diffusion coefficient $\kappa_{\perp}$ from $\kappa_{\parallel}$ using the Non-Linear Guiding Center (NLGC) Theory \citep{shalchi2010analytic},
\begin{equation}\label{eq:kappa_perp}
\kappa_{\perp} = \left [ \frac{\sqrt{3}}{3} va^{2}\pi C(\nu) \frac{\delta B^{2}_{2D}}{B^2_{0}} l_{2D}\right ]^{2/3}\kappa_{\parallel}^{1/3},
\end{equation}
where the square of the turbulence magnetic field follows a radial dependence of $\delta B^{2}\sim r^{\gamma}$. We assume the ambient turbulence level $\delta B^{2}/B^{2}$ to be $0.15$ at 1 au. Recently, \citet{Hu+etal+2018,Ding+2020} considered the radial dependence of the bendover scales $l_{slab}$ and $l_{2D}$ to have a form of $r^{a_{\rm slab}}$ and $r^{a_{\rm 2D}}$. With Eq.~(\ref{eq:kappa_para}) and ~(\ref{eq:kappa_perp}),  one obtains,
\begin{equation}\label{eq:radial dependence}
\begin{aligned}
&\kappa_{\parallel} \sim v^{\frac{4}{3}} B^{\frac{5}{3}} r^{\frac{2}{3} a_{\rm slab}-\gamma}; \\
&\kappa_{\perp} \sim  v^{\frac{10}{9}} B^{-\frac{7}{9}} r^{\frac{\gamma}{3}+\frac{2a_{\rm slab}}{9}+\frac{2a_{\rm 2D}}{3}}.
\end{aligned}
\end{equation}
In this work, we assume $\gamma=-3.5$, $a_{\rm slab}=1.0$ and $a_{\rm 2D}=1.0$, the ratio of $\kappa_{\perp}$ to $\kappa_{\parallel}$ therefore has a form of $\kappa_{\perp}/\kappa_{\parallel } \sim v^{-2/9} B^{-22/9} r^{-40/9}$. At 1 au, a reference value of $\kappa_{\perp}/\kappa_{\parallel }=0.03$ for $10$ MeV proton is chosen.

\subsection{Coupling}
In this work, EUHFORIA is used to describe both the data-driven background solar wind and the CME from $0.1$ au to $2$ au. The CME-driven shock is identified and the calculated shock parameters along the 3D shock surface at different times are imported to the iPATH model. Note that we use the full information of the 3D shock in the iPATH model. The accelerated particle spectrum at the shock front is computed based on the shock parameters. A 3D shell model behind the shock is constructed based on the modelled shock fronts from EUHFORIA, and these shells propagate out with the shock. The iPATH calculates the convection and diffusion of the accelerated particles among multiple shells. When particles escape upstream of the shock, they propagate along the magnetic field line. For simplicity, we assume the magnetic field upstream of the shock is given by the Parker field in the transport module. The Parker field line through each spacecraft is corresponding to the simulated solar wind speed at the flare onset time. Using a backward stochastic differential equation method, we follow the propagation of these energetic protons with cross-field diffusion included in the solar wind. We note that the setup of diffusion coefficients are the same for different observers since we do not utilize the information on the background solar wind from EUHFORIA. Finally, the modelled time intensity profiles are obtained at PSP, STA, Earth and SolO and compared to observations.

In short, we use full information from EUHFORIA to model the particle acceleration at the shock but we still use the Parker field to study the particle transport in the iPATH model. We do not use the magnetic field lines from EUHFORIA due to the limits of the current transport module in iPATH. However, we note that in the realistic solar wind, multiple fast and slow streams exist and their presence will complicate the interplanetary magnetic field. Furthermore, preceding CMEs can disturb the interplanetary environment and lead to a distorted interplanetary magnetic field. Taking into account the presence of solar wind streams with varying speeds,  the PARADISE model \citep{wijsen2019modelling} solved the focused transport equation with a data-driven solar wind generated from the EUHFORIA. This approach has shown a remarkable capacity to capture detailed features of CIR events and ESP events \citep{wijsen2021observation,wijsen2021self}. Extending it to a non-Parker field in the iPATH model will be pursued in future work.

\section{Results}\label{sec:results}

\subsection{Shock properties}
\begin{figure*}
\includegraphics[width=16.9cm]{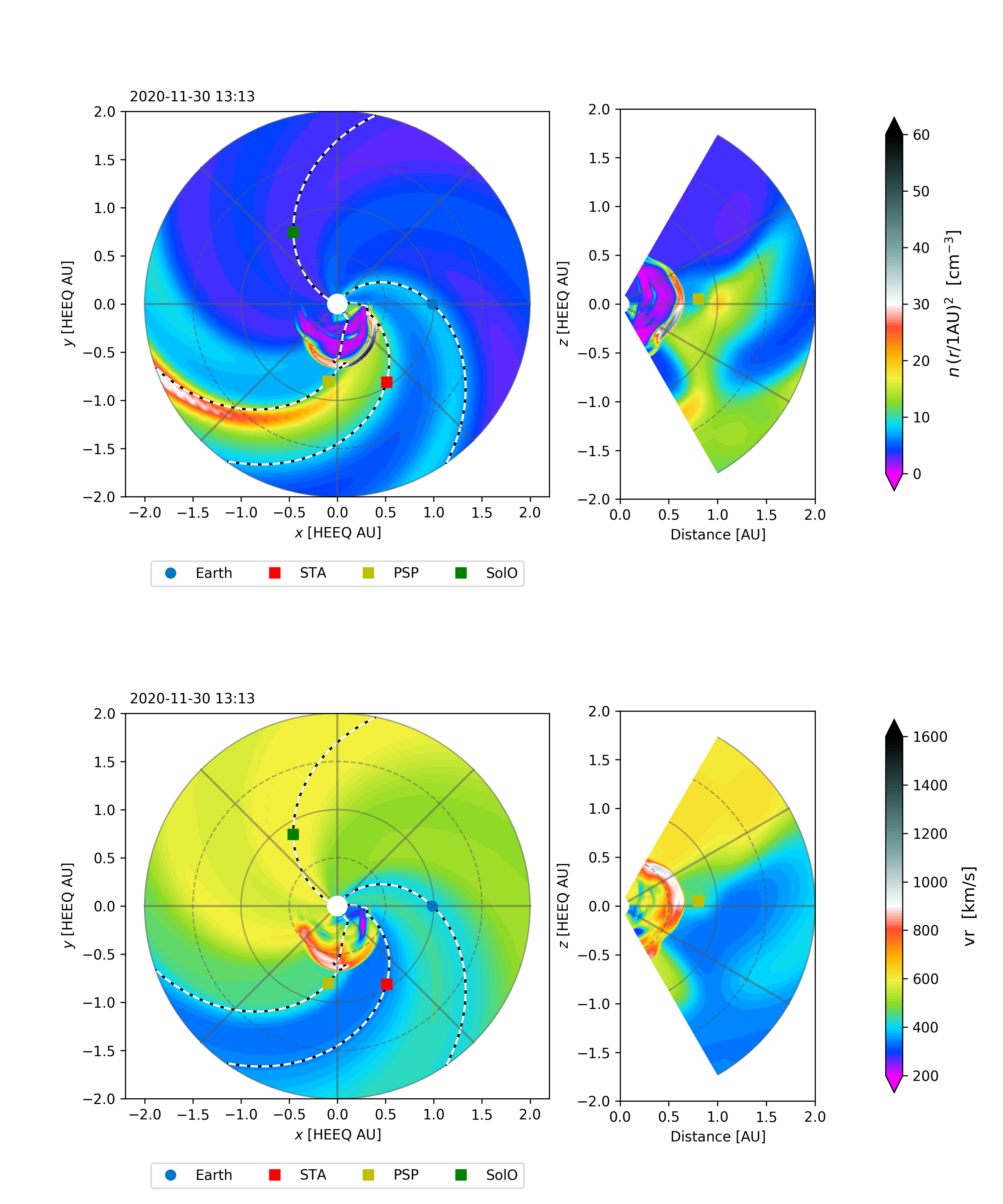}
\caption{Equatorial and meridional snapshots of the modelled scaled number density (top row) and the radial speed (bottom row) in EUHFORIA. The left column  represents the equatorial plane, while the right column shows the meridional plane that includes PSP. The dashed curves show the interplanetary magnetic field lines  corresponding to Earth, STA, PSP and SolO.}
\label{fig:euhforia}
\end{figure*}

Figure~\ref{fig:euhforia} shows snapshots of the scaled number density and the radial speed from the EUHFORIA simulations at 13:13 UT, on November 30 2020. 
Snapshots are presented in the Heliocentric Earth Equatorial (HEEQ) coordinate system. The locations of four spacecraft and the corresponding simulated and observed solar wind speeds at the spacecraft locations 
are listed in Table~\ref{table2-sc_loc}. From the table we can see that at STA, the simulated and observed solar wind speeds are similar; at Earth,  the simulated solar wind speed is faster than the observation. Solar wind measurements from PSP and SolO are not available and are shown by N/A in the table.

In the iPATH model, the CME-driven shock is modelled by a hydrodynamic disturbance at the inner boundary and there is no explicit description of the CME itself. Consequently, the magnetic pressure of the CME flux rope is ignored in the iPATH model. 
However, the expansion of CMEs, hence the properties of the CME-driven shock, are heavily affected by the internal magnetic field structure of CMEs \citep{scolini2019observation}. 
In this work, the CME is initialized using the LFFS model and the input parameters of the LFFS model are listed in Table~\ref{table1-lffs}. The center of the CME is -$80^{\circ}$ in longitude. In the equatorial snapshot of the radial speed, the propagation speed at the eastern portion of the CME is much faster than that of the western portion. Such an asymmetric expansion of CME is due to variations of upstream solar wind conditions, that is, the CME propagates faster in fast streams than in slow streams. This is also evident from the meridional slices, where a faster expansion of the CME towards higher northern latitudes is observed (see the upper right panel of Fig.~\ref{fig:euhforia}). These snapshots reveal the fact that  background solar wind structures can considerably affect the propagation and expansion of a CME. Such an asymmetric expansion further impacts the shock speed distribution and shock geometry along the shock surface. In Fig.~\ref{fig:euhforia}, the locations of Earth, STA, PSP and SolO spacecraft are marked by colored circles and squares. The white dashed lines are magnetic field lines connecting to these locations. On November 30 2020 at 12:13 UT,
PSP is well connected to the nose of the shock and STA is marginally  connected to the flank of the shock. In contrast, SolO and Earth are not connected to the shock in this simulation. From the in-situ measurements of SEPs, \cite{kollhoff2021first} used Velocity Dispersion Analysis (VDA) and Time Shift Analysis (TSA) methods to calculate the particle release times, which yields a solar release time of 13:15 UT at PSP and of 14:47 UT at STA, for $\sim 16$ MeV protons. However,  from the SolO observation, they estimated the release time of $\sim 16$ MeV protons to be 13:53 UT on November 29 2020, which is earlier than that at STA. This suggests that SolO has a better magnetic connection to the shock than STA early in this event, contradicting to the simulation results where SolO does not connect to the shock. We further discuss it below in the SEP simulation results of SolO.

\begin{table}
\caption{\label{table2-sc_loc}Location of the spacecraft  }
\centering
\begin{tabular}{lccccc}
\hline\hline
S/C & r(au) &Lon.\tablefootmark{a}  & Lat.\tablefootmark{a}   & $U_{sim}$\tablefootmark{b} (km/s)  & $U_{obs}$\tablefootmark{c}(km/s)  \\ 
\hline
PSP & 0.81 & -96$^{\circ}$ & 3$^{\circ}$ & 358 & N/A\tablefootmark{d}\\
STA & 0.96 & -57$^{\circ}$&  6$^{\circ}$ & 368 & 361\\
Earth &0.99 & 0$^{\circ}$&  1$^{\circ}$ & 436 & 358\\
SolO & 0.88 &122$^{\circ}$& -5$^{\circ}$ & 596 & N/A\\

\hline
\end{tabular}\\
\tablefoottext{a}{Coordinates are given in the HEEQ coordinates.}\\
\tablefoottext{b}{$U_{sim}$ is the solar wind speed given from EUHFORIA on 29/11/2020 at 13:00 UT }\\
\tablefoottext{c}{$U_{obs}$ is the solar wind speed from in-situ measurements on 29/11/2020 at 13:00 UT.}\\
\tablefoottext{d}{ N/A refers to no available data at spacecraft during this event.}
\end{table}

\begin{figure}
   \centering
   \includegraphics[width=\hsize]{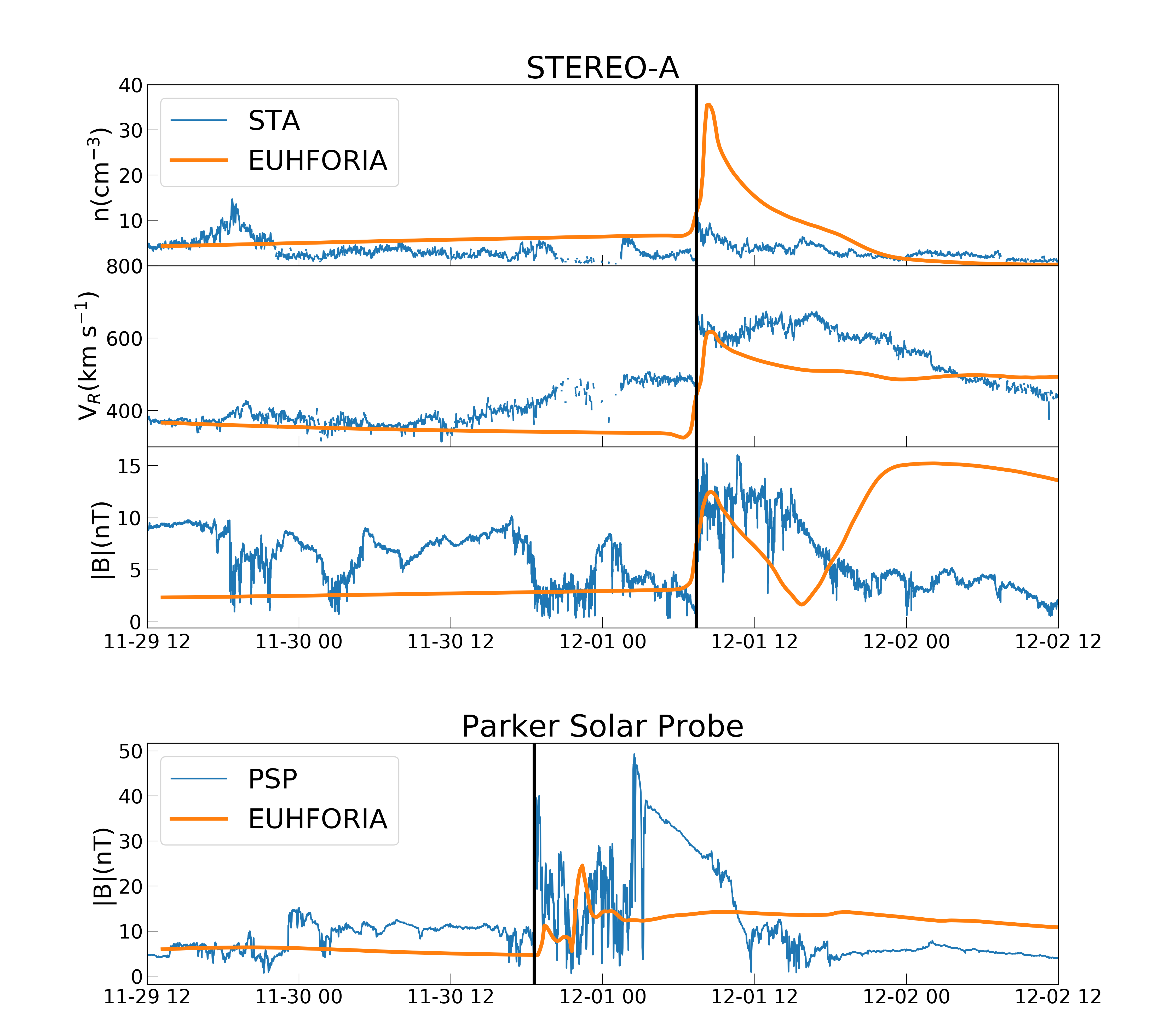}  
\caption{Time series of the modelled plasma and magnetic field parameters at STA and PSP. The blue lines indicate the in-situ observations and the orange lines show the modelled results of EUHFORIA. The vertical solid lines indicate the shock arrival time at STEREO-A and PSP (See text for details).  }
\label{fig:time-profile-sw}
\end{figure}

Figure~\ref{fig:time-profile-sw} shows simulated time profiles of the solar wind at STA and PSP from 2020-11-29 12:00 UT  to 2020-12-02 12:00 UT, compared to in-situ measurements. The upper panel shows proton number density, radial speed and magnetic field magnitude at STA. The yellow lines represent the modelled results in EUHFORIA and the blue lines show in-situ measurements. The vertical solid line indicates the shock arrival at STA, which occurred on December 1 2020 at 07:23 UT. 
Prior to the shock passage, the observed and simulated proton number density and magnetic field magnitude are similar, but there is a high speed stream (HSS) around 12-01 from the observation which is not captured by the simulation. After the shock passage, the simulated and observed speed and magnetic field magnitude are similar, but the simulated number density is significantly higher than the observation. This is related to the fact that the simulation does not contain a HSS upstream of the shock. In the shock frame, a HSS upstream of the shock means a smaller flux toward the shock, and therefore leads to a smaller density downstream of the shock.  
Since no solar wind measurements are available from PSP, we only compare the magnitude of the magnetic field between the model and the measurement in the lower panel of Fig.~\ref{fig:time-profile-sw}. The shock reached PSP on November 30 2020 at 18:35 UT, which is indicated by the vertical line. This is similar to the simulated shock arrival time. We note that from the observation, a clear shock sheath can be seen after the shock arrival in both spacecraft. In comparison, the shock complex structures are not clear in the EUHFORIA simulation.  This is not an issue since  we do not discuss the shock complex in this work. We use the shock arrival time at PSP and STA to constrain the initial boundary conditions of the CME.

\begin{figure*}
   \centering
\includegraphics[width=16.9cm]{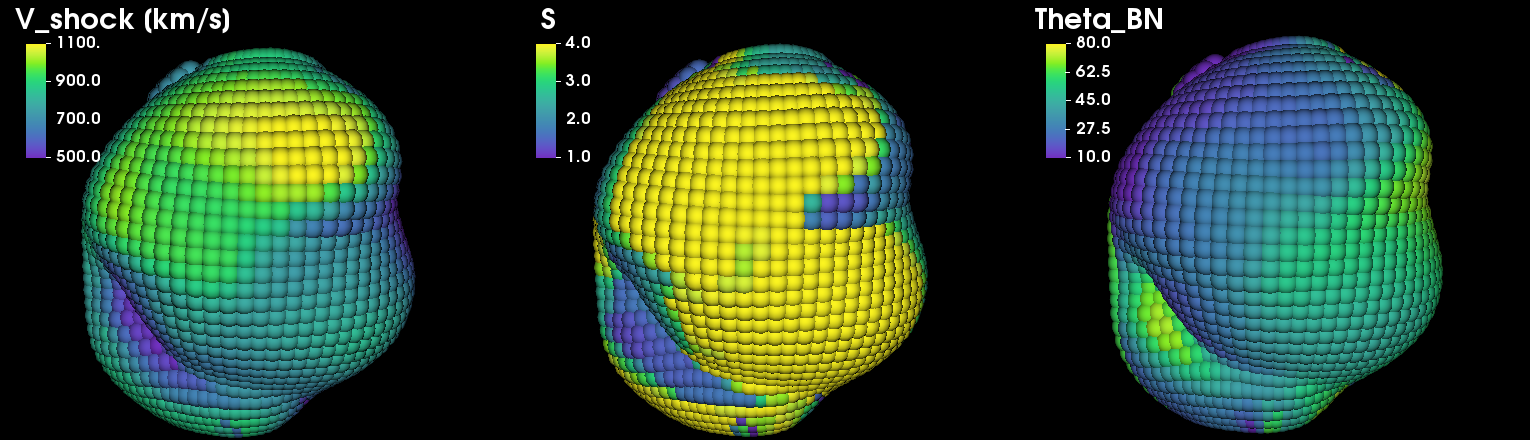}  
\caption{Shock parameters along shock surface at $t=30$ hours, showing the shock speed $V_{\rm shock}$, the compression ratio $S$ and the shock obliquity $\theta_{\rm BN}$. }
\label{fig:shock_paras_3d}
\end{figure*}

In our simulation, we identify the shock location and calculate shock parameters along the shock surface every 2 hours as the shock propagates out. Figure~\ref{fig:shock_paras_3d} shows the shock speed, the compression ratio and the shock obliquity along the shock surface on  November 30 2020 at 21:13 UT. In this 3D paradigm, it can be seen that the shape of the shock is clearly modulated by the solar wind structure. In the southern hemisphere (bottom part of the shock), a dip of the shock can be seen from all three panels. It results from the interaction between the CME and a corotating interaction region (CIR). This interaction significantly changes the geometry of the shock, as well as the shock strength. It illustrates that the interaction between CIR and CME can be essential to understand shock acceleration and transport. Asymmetric coronal shocks and their corresponding shock acceleration have been examined by \cite{Li2021,jin2018probing,jin2022assessing}. These authors suggested that the asymmetric expansion of coronal shocks is essential in understanding the characteristics of SEP events. We remark that the combination of EUHFORIA and the iPATH model is a powerful approach to model the SEP events in the inner heliosphere. In the left panel, the shock region with the highest shock speed appears in the northern hemisphere(the upper portion of the shock). This is due to the expansion of the CME in a fast stream as shown in the meridional slices of Fig.~\ref{fig:euhforia}. The middle panel depicts the compression ratio along the shock surface. The compression ratio at the shock nose is the highest, close to 4, whereas it decreases from the shock nose to the flank of the shock. In DSA, the compression ratio decides the spectral index of the source spectra. That is, a higher compression ratio results in a harder spectrum. It is worth noting that the compression ratio is lowest at the dip, indicating a weaker acceleration efficiency in this portion of the shock. In addition to the compression ratio, the shock geometry plays a critical role in deciding the acceleration efficiency of the shock \citep{Li+2012}. The right panel shows that the shock obliquity angle gradually increases from the eastern flank (left side) to the western flank (right side), that is, the shock changes from being quasi-parallel at the eastern flank to being quasi-perpendicular at the western flank.

\begin{figure*}
   \centering
\includegraphics[width=\hsize]{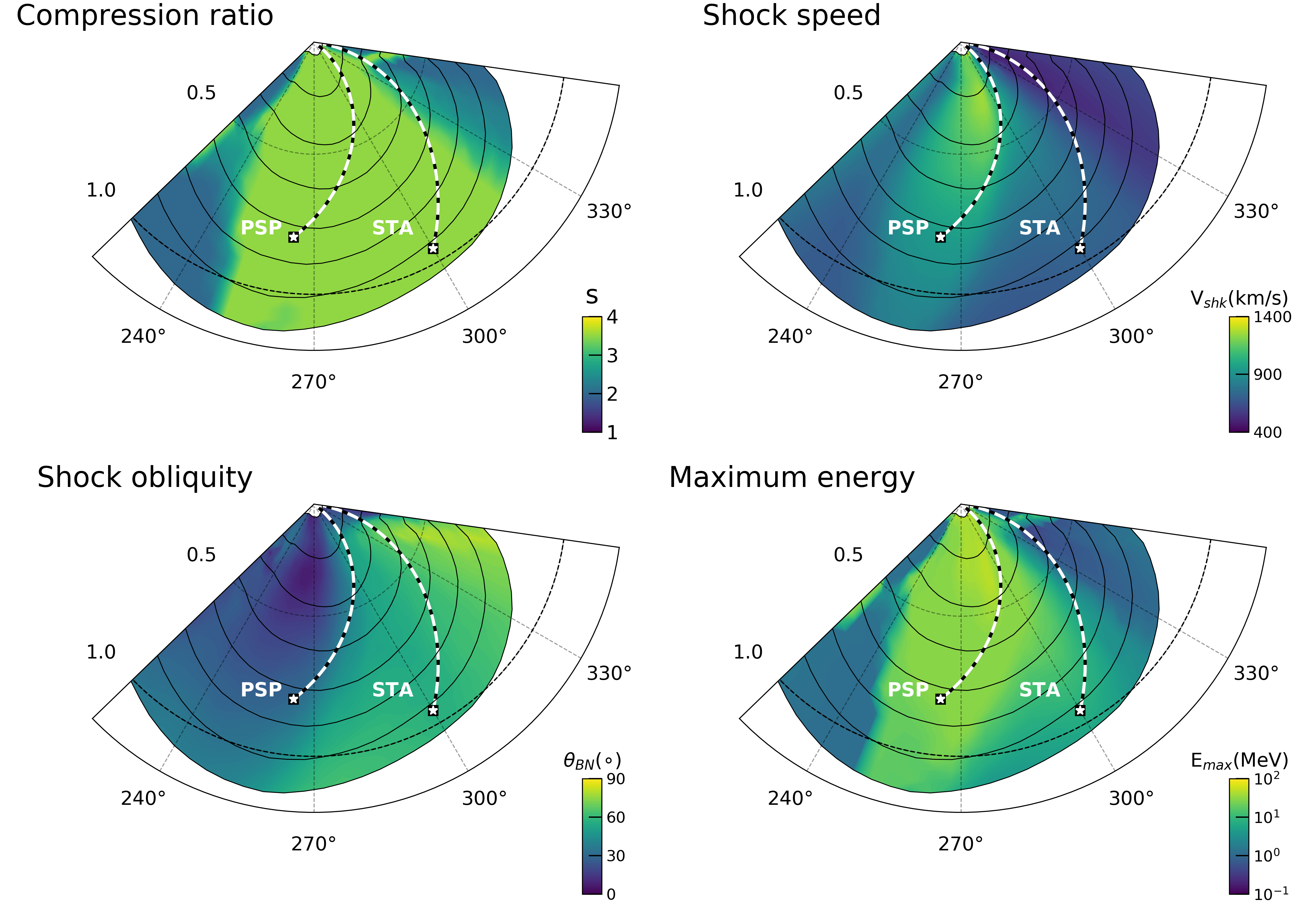}  
\caption{The evolution of shock location and shock parameters in the equatorial plane from $0.1$ au to $1.2$ au. The black solid curves show the shock front at different time steps. The color schemes are for different shock parameters along the shock front. The white dashed curves signal the Parker magnetic field lines passing through PSP and STA. }
\label{fig:shock_ecliptic}
\end{figure*}

To clearly demonstrate the history of shock acceleration, Figure~\ref{fig:shock_ecliptic} plots the time evolution of shock parameters in the equatorial plane. The shock parameters shown in the four panels are the shock compression ratio, the shock speed, the shock obliquity and the maximum proton energy. In each panel, black curves are shock fronts at a series of times from $0.1$ au to $1.2$ au and the color scheme indicates the magnitude of the corresponding shock parameters along the shock front. We assume that PSP and STA are located in the solar equatorial plane as their latitudes are $3^{\circ}$ and $6^{\circ}$ respectively. White dashed curves are the Parker field lines that pass through PSP and STA at the beginning of the event, respectively. We assume the spacecraft do not move significantly during the event so that these Parker field lines do not change over the course of the event.
Earth and SolO are not connected to the shock in the simulation so they are not shown in this figure. As seen from the panel of the compression ratio, PSP always connects to the high compression region ($>3$) along the magnetic field line, whereas STA connects to the edge of shock with a lower compression ratio early on and later connects to a stronger compression ratio region when the shock is beyond $0.5$ au. The upper right panel depicts the history of shock speed. From the figure we can see that, before the shock reaches $0.5$ au, the PSP connects to a shock region with a relatively high shock speed ($V_{\text{shock}} > 1000$ km/s) along the magnetic field line. In contrast, STA connects to the shock flank with a low shock speed ($V_{\text{shock}}< 600$ km/s). Beyond $0.5$ au, the shock speed for the PSP connection region drops somewhat and the shock speed for the STA connection region increases slightly. However, the PSP-connected region is still faster than the STA-connected region.  
The bottom left panel shows the shock obliquity in the solar equatorial plane. We can see that the eastern flank (to the left) of the shock always maintains a quasi-parallel geometry but the western flank (to the right) varies from a quasi-parallel geometry to a quasi-perpendicular geometry. 
From the evolution of the magnetic connections at PSP and STA, we can conclude that PSP observes a stronger SEP event since the shock region connected to PSP maintains a high compression ratio, high shock speed and a quasi-parallel geometry throughout the event. Using Eq.~(\ref{eq:dynamic time}), we calculate the maximum proton energy at the shock front based on the shock parameters, shown in the bottom right panel. The distribution of the maximum proton energy reflects the efficiency of shock acceleration. We point out that because STA connects to the weak shock flank ($< 0.5$ au) early on, and later to a stronger shock region, the maximum proton energy at the shock front (along the STA-connected field line) increases over time and can reach tens of MeV when the shock is beyond $0.5$ au. We note that the history of shock acceleration directly impacts the time intensity profiles observed by different spacecraft.

\subsection{Time intensity profiles}
\begin{figure}
   \centering
   \includegraphics[width=\hsize]{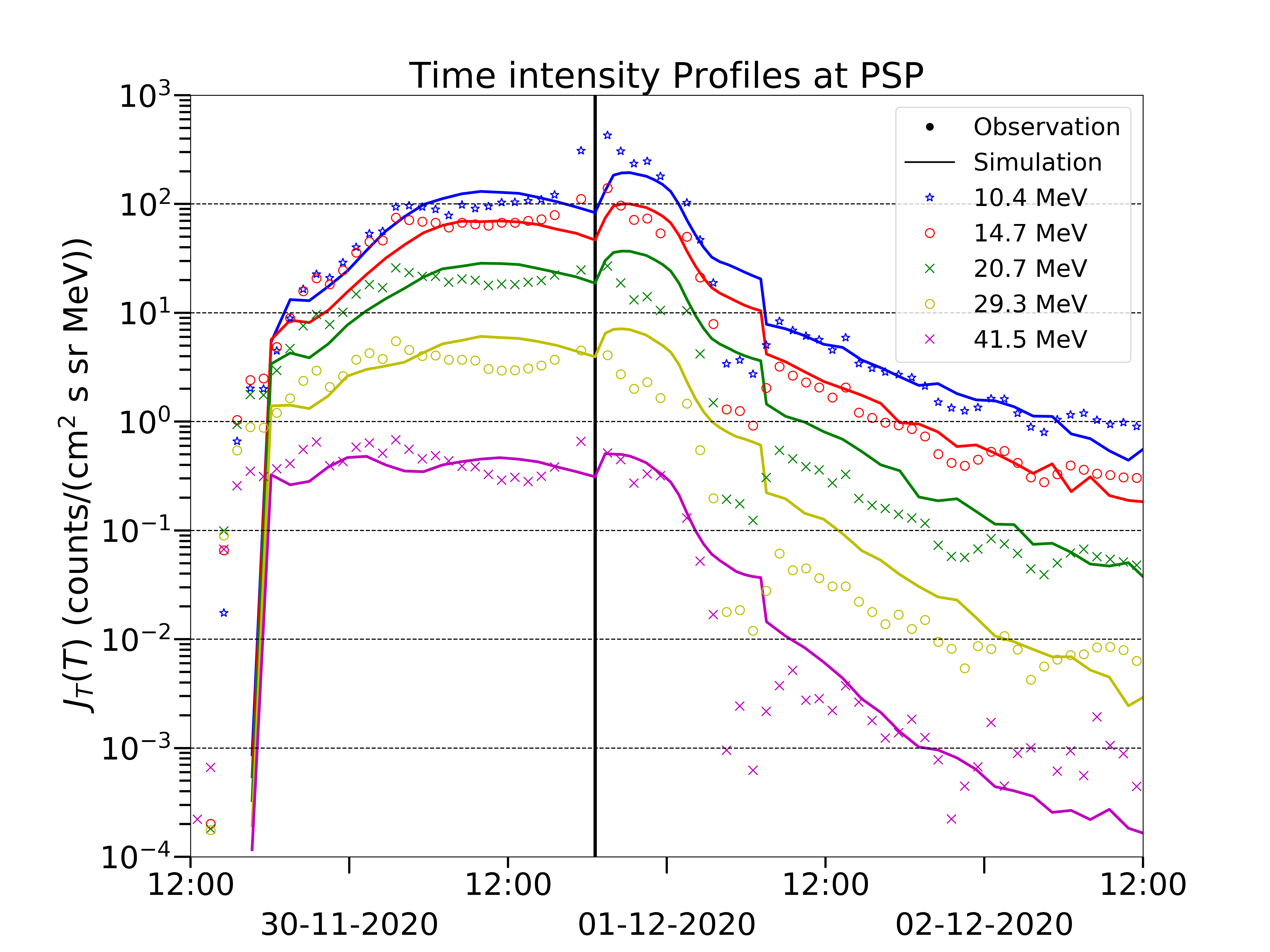}  
\caption{Time intensity profiles  from the PSP observation (symbols) and the model calculation (solid lines).  Five energy channels are shown from the observation and are same in the simulation. The vertical line marks the shock arrival at PSP in the observation.}
\label{fig:psp}
\end{figure}

After computing the particle spectrum at the shock front and obtaining the escaped particle spectra at a series of times, we follow the  transport of these particles and obtain the time intensity profiles at four spacecraft. In this work, we focus on the comparison of the observed and modelled time intensity profiles for proton energies greater than 10 MeV, from 2020-11-29 12:00 UT to 2020-12-02 12:00 UT.  
Figure~\ref{fig:psp} shows the observed (points) and the modelled (lines) proton time profiles at PSP. The in-situ measurements come from the High Energy Telescope A (HETA) of the Energetic Particle Instrument-High (EPI-Hi; \citet{Wiedenbeck2017ICRC...35...16W}) as part of the Integrated Science Investigation of the Sun (IS$\odot$IS; \citet{mccomas2016integrated}) on board PSP. Five energy channels: $10.4$ MeV, $14.7$ MeV, $20.7$ MeV, $29.3$ MeV, and $41.5$ MeV are shown. The time resolution of data points in the observation is $1$ hour. The vertical line indicates the shock arrival time in the observation. The modelled time intensity profiles successfully capture several important features including the gradual increase and plateau-like phase, the energetic storm particle (ESP) phase and the decay phase. Before the shock arrival, time intensity profiles of low proton energies ($< 30$ MeV) show a gradual enhancement before 2020-11-30 6:00 UT and then a plateau period until the shock arrival. In contrast, the time intensity profile of $41.5$ MeV has a prompt increase at the beginning of the event which is followed by a plateau until the shock reaches PSP. 
This indicates that $> 40$ MeV protons are accelerated near the Sun early on and are released from the shock complex during an extended period. In contrast, the injection of  $< 30$ MeV protons continues at the shock until around 2020-11-30 6:00 UT. This continuation release leads to the gradual enhancement observed at PSP. 
The ESP phase follows the shock passage. Observation of the ESP phase is the most prominent for the $10.4$ MeV and $14.7$ MeV channels. From the figure we can see that  both the modelled enhancements and the modelled duration of the ESP phase for these two energies are similar to the observation. In comparison, the enhancements of the ESP phase for the $20.7$ MeV and $29.3$ MeV protons in the model are larger than the observations. This suggests that the modelled source spectrum in the shock complex is harder than that in the observation. We note that the modelled duration of the ESP phase is decided by the shell structure as explained in Sect.~\ref{sec:model}. Among the shells behind the shock, accelerated particles can diffuse and convect. After the shock passage,
the distribution function in these shells maps to the ESP phase time profiles as observed by the spacecraft. We note that there are different models of the ESP phase. Recently, \citet{wijsen2021self} modelled the ESP event of 14 July 2012 using the PARADISE transport model and obtained a reasonable agreement with the observation. They prescribe particle parallel mean free path near the shock and inject $50$ keV protons at the shock. The proton intensities below $\sim$ 1 MeV are well reproduced both upstream and downstream during the ESP phase. However, this approach cannot capture higher energy particles in the ESP phase, which are likely accelerated at an earlier time instead of being locally accelerated at $1$ au. In comparison, the treatment of a shell structure in the iPATH model naturally contains pre-accelerated particles within the shock complex. 
We note that the duration of the shock complex can be used to examine how good the shell model is in capturing the trapping of energetic particles.

\begin{figure}
   \centering
   \includegraphics[width=\hsize]{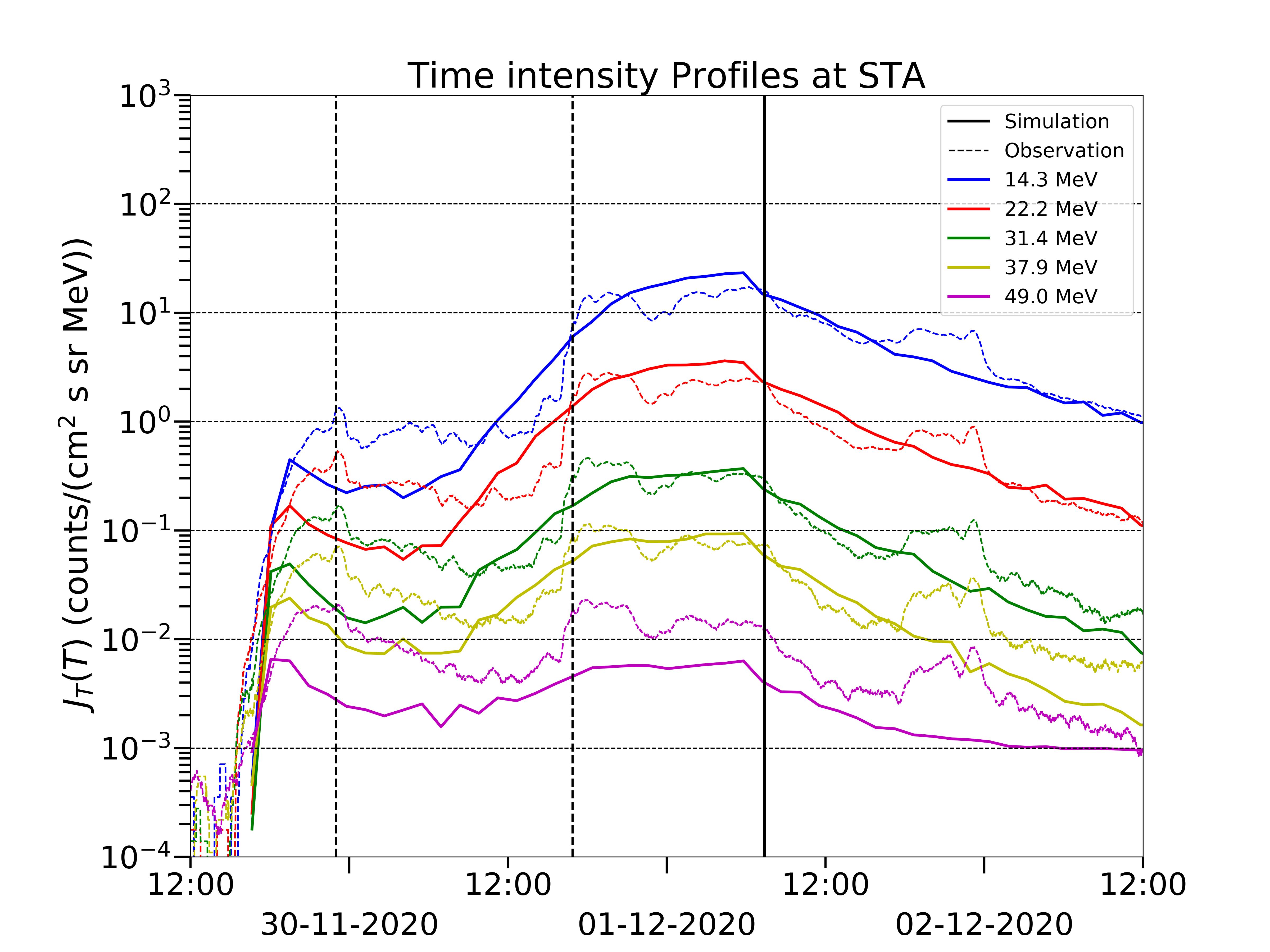}  
\caption{Time intensity profiles from the STA observation (dashed lines) and the model calculation (solid lines). The solid vertical line marks the shock arrival at the STA in the observation and the dashed vertical lines indicate the heliospheric plasma sheets.}
\label{fig:sta}
\end{figure}

We next examine the time intensity profiles at STA. Figure~\ref{fig:sta} shows the modelled (solid lines) and the observed (dashed lines) time intensity profiles for five energy channels, labelled in the upper left corner. The two vertical dashed lines indicate two heliospheric plasma sheets (HPS) (See Figure 2 of \citet{kollhoff2021first}) and the solid vertical line indicates the shock arrival at STA. The five energy channels in the model are the geometric means of $5$ energy bins from the High Energy Telescope (HET;\citet{von2008high}) on board STA. The observed proton intensity shows a fast enhancement at the beginning and then a second abrupt enhancement which occurs at around 2020-11-30 17:00 UT.  \cite{kollhoff2021first} pointed out that the proton intensities between the two heliospheric plasma sheets were modulated by a local solar wind structure with a northward magnetic field. When STA left the local solar wind structure and entered the second HPS, an abrupt increase of the proton intensity occurred. A recent study by \citet{waterfall2022modeling} showed that the heliospheric current sheet (HCS) plays a role in promoting large SEP events due to the current shift drift.
In a previous modelling effort, \citet{odstrvcil1996propagation} has considered a case where an interplanetary shock propagates along the HPS. They found that a dip is formed at the shock front. This implies that the presence of HPS can affect the shock parameters and shock geometry. Since in our simulation, no HPS is identified, the influences of the HPS are absent in this work. Nevertheless, the modelled results do show a two-step enhancement resulting from the  evolution of the shock properties. As shown in Fig.~\ref{fig:shock_ecliptic}, STA connects to the edge of the western shock flank with a low shock speed early on. Consequently, the proton intensity observed at STA was lower than that at PSP. Beyond $0.5$ au, STA connects to a stronger portion of the shock where the shock speed and the compression ratio are both larger. This lead to a second enhancement of the intensity, which starts at around 7:00 UT. After the shock arrival, time intensity profiles in the observation and the simulation show similar decay behavior, especially at low energies. We suggest that the two-step enhancement of proton intensity results from the history of extended particle acceleration at the shock, which depends on the evolution of the shock parameters. We note that the second enhancement from the observation is rather gradual, lasting $\sim$ 6 hours, and the sudden enhancement around 18:00 UT is at the end of this gradual enhancement. While the HPS should be responsible for this sudden change at 18:00 UT, our model addresses the more gradual enhancement.  We do emphasize that HPS can modulate the transport of energetic particles, as the recent observations of Jovian electrons by PSP suggested \citep{Mitchell2021}.  
However,
to understand the role of HPS in this event,  a comprehensive modelling work on the interaction between shock and HPS is necessary, which is out of the scope of this work.
\begin{figure}
   \centering
   \includegraphics[width=\hsize]{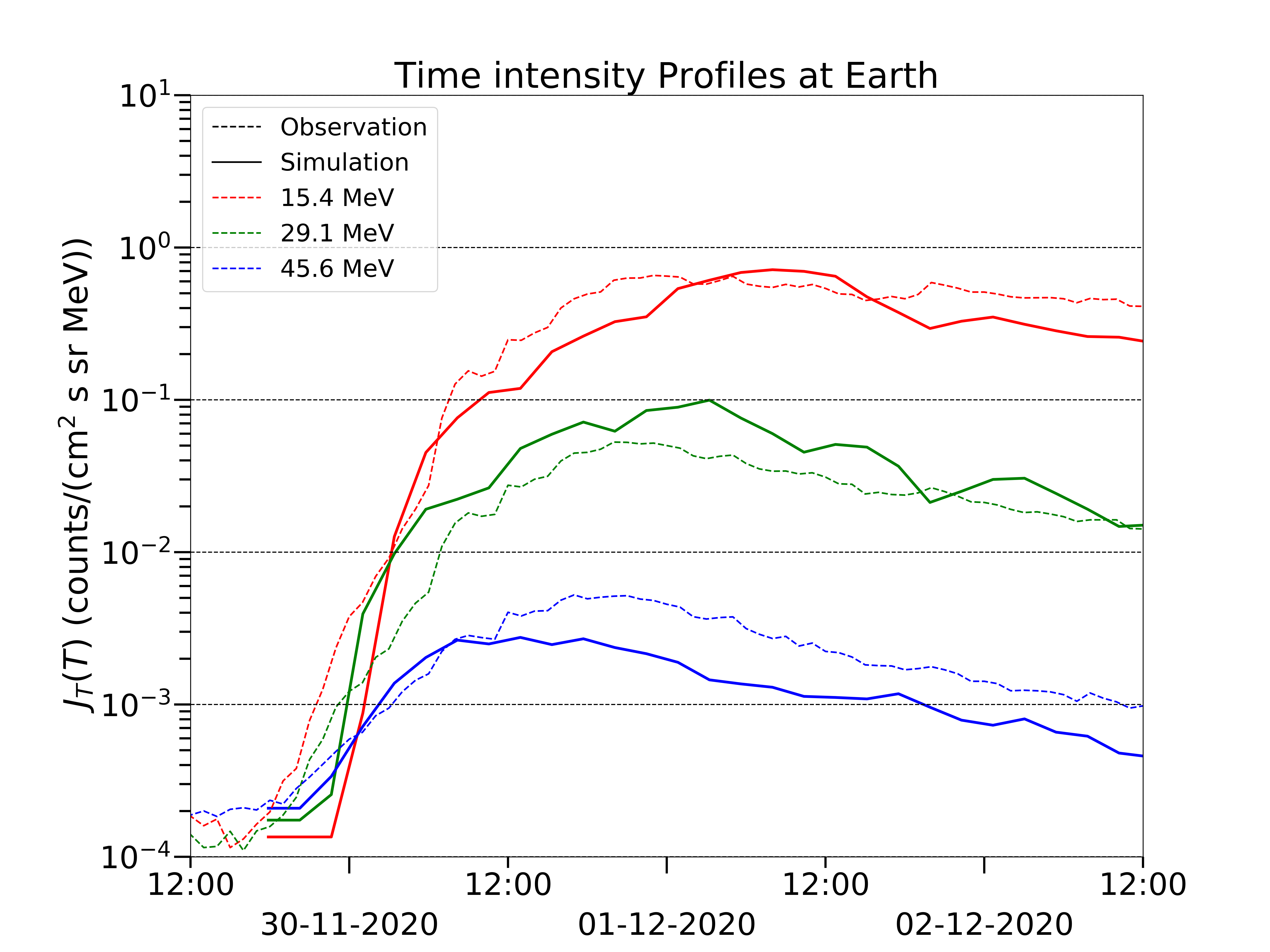}  
\caption{Time intensity profiles  from the Earth observation (dashed lines) and the model calculation (solid lines).}
\label{fig:earth}
\end{figure}
 
We next examine time intensity profiles of protons as observed by SOHO/ERNE \citep{torsti1995energetic}. Protons from three energy channels: $15.4$ MeV, $29.1$ MeV, $45.6$ MeV are shown in Fig.~\ref{fig:earth} for both simulations and observations. The observation data is shown by the dashed lines and the solid lines are for the simulation. We note that the Earth’s magnetic footpoint is approximately $157$ degrees west of the flare. As shown in Fig.~\ref{fig:euhforia}, the shock simulated with EUHFORIA only extends to around 0$^{\circ}$ in longitude, so that Earth is not connected to the shock until the shock  passes 1 au. There is no plasma signature of shock arrival from near-Earth spacecraft. Therefore, particles observed at the Earth early on likely propagate to the magnetic field line that connects to Earth by cross-field diffusion. The diffusion process both parallel and perpendicular to the magnetic field depends on the turbulence in the solar wind. One important aspect of the diffusion process is the radial evolution of magnetic turbulence. From the newest observations of PSP, some recent observational and theoretical studies suggest the correlation lengths of quasi-2D and slab turbulence have a radial dependence of $r^{a}$ where $\alpha$ is close to $1$ \citep{Adhikari2020,Chen2020}. We assume $a=1$ as a reference value in this work. We note that the peak intensity observed at the Earth directly relates to the strength of cross-field diffusion, that is, the stronger cross-field diffusion, leading to the higher peak intensity. Hence, the choice of the perpendicular diffusion coefficient is bounded by the observations. Furthermore, the enhancement of proton intensity prior to the peak intensity is determined by the strength of both the perpendicular and parallel diffusion. Under these constraints, in this work, we utilize the diffusion coefficient form introduced in Sect.~\ref{sec:model}.     
Comparing to the observation, the model shows a faster and larger enhancement of proton intensity for the rising phase, but for the decay phase the model result is remarkably similar to the observation. Since the inner boundary of the EUHFORIA is at $0.1$ au, particle acceleration and transport below $0.1$ au are not considered in this work. This may explain the delayed enhancement at the beginning of the event. The intensity of $45.6$ MeV protons is slightly lower than that in the observation, indicating the acceleration of high energy protons is not adequate in the model. Since high energy protons are mainly accelerated near the sun (below $0.1$ au), this work indeed missed part of high energy protons.
The comparison shows that a proper choice of perpendicular diffusion and parallel diffusion is crucial to model the SEP event observed at the Earth. In particular, perpendicular diffusion is the key to understand SEP events for those observers without a good magnetic connection to the shock.

\begin{figure}
   \centering
   \includegraphics[width=\hsize]{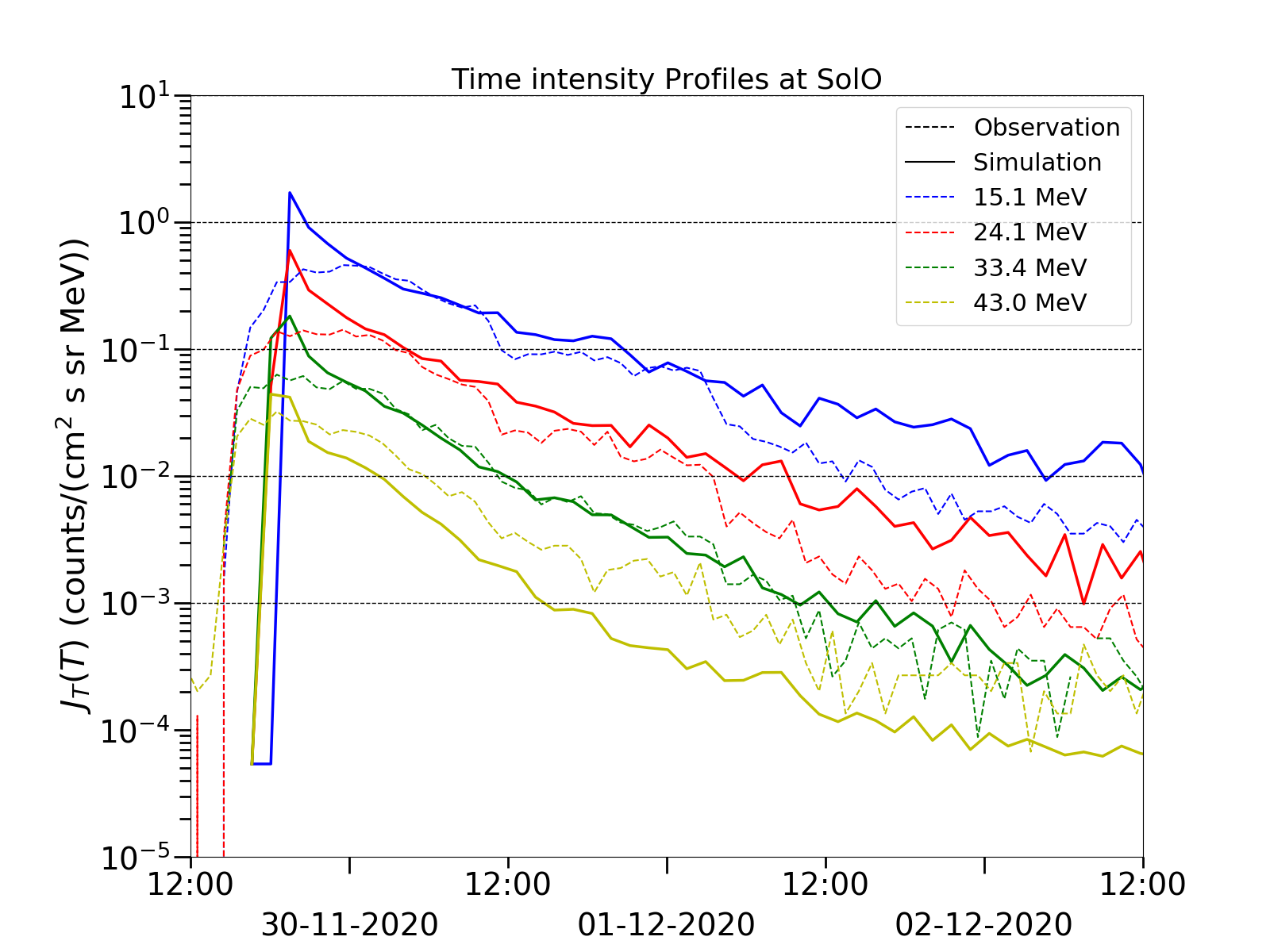}  
\caption{Time intensity profiles  from the SolO observation (dashed lines) and the model calculation (solid lines). We note that the model results are obtained with an assumption of Parker magnetic field related to a solar wind speed of $160$ km/s. See text for details. }
\label{fig:so}
\end{figure}

Finally, we consider the time intensity profiles of protons at SolO. As viewed from SolO, the event is a backside event. However, the time profile still shows a prominent fast enhancement after the onset of the flare. \citet{kollhoff2021first} 
pointed out that the proton onset times at PSP and SolO are similar. We note that PSP has a good magnetic connection during this event as shown in Fig.~\ref{fig:shock_ecliptic}. A similar onset time at SolO is only possible if the SolO has a good magnetic connection to the CME-driven shock early on. However, our simulation shows that the SolO does not connect to the shock during the entire event. It is possible to have the SolO magnetically connected to the shock if the solar wind speed is low. Since there is no available solar wind plasma data of SolO during the event, \citet{palmerio2022cmes} estimated the solar wind speed based on the data from Magnetometer and the Radio and Plasma Waves instrument on board SolO using a deHoffmann–Teller analysis. These authors showed that there is likely a transition of solar wind speed from a slow speed stream to a high speed stream at the onset of SEPs  in their Figure A6 \citep{palmerio2022cmes}. This indicates that SolO may be associated with a transient structure at the onset of this event. Similarly, \citet{kouloumvakos2022first} marked a clear SIR structure passing through SolO at the onset of this event using the ENLIL simulation. In Figure~7 of \citet{kouloumvakos2022first}, the structure of SIR2 is largely curved. Based on these studies,  a reasonable scenario is the following:  at the onset of this event, SolO is passing through a SIR structure and magnetic field lines in the SIR were distorted from the nominal Parker field lines such that the footpoint of the field line that connects to SolO is very close to the flare. However, EUHFORIA does not capture this transient SIR structure, partly because it does not include a real-time inner boundary condition. To mimic the effect of magnetic field lines with a large curvature, here we assume SolO is at a Parker spiral magnetic field with a solar wind speed of about $160$ km/s early in this event. This workaround is not to be taken as physically, but to yield a smaller longitudinal separation $\sim 20^{\circ}$ between the magnetic footpoint and the flare location at the inner boundary. Since the peak proton intensity decreases with increasing offsets between footpoints and the flare location \citep{Ding2022}, the choice of a $160$ km/s solar wind speed is to fit the peak intensity to observations. Although this assumption is simple, the modelled results fit the observed time intensity profiles very well.  Figure~\ref{fig:so} shows the comparison between the observed and simulated time intensity profiles at SolO. Four energy bins: $15.1$ MeV, $24.1$ MeV, $33.4$ MeV and $43.0$ MeV are selected from the High Energy Telescope (HET) in the Energetic Particle Detector (EPD; \citet{rodriguez2020energetic}) onboard SolO. HET has four telescopes with sunward facing, anti-sunward facing, northward facing and southward facing. The averaged-directional data are plotted by the dashed lines, and the solid lines represent the simulation results. The modelled peak intensities for all four energy channels are slightly higher than those from the observation, but the decay phases agree remarkably. This comparison shows that the observation at SolO can be explained by an extremely  curved magnetic field which has a reasonable magnetic connection to the CME-driven shock. Such  an extremely curved magnetic field can be due to the presence of a SIR.

\subsection{Role of cross-field diffusion}
\begin{figure*}
   \centering
   \includegraphics[width=\hsize]{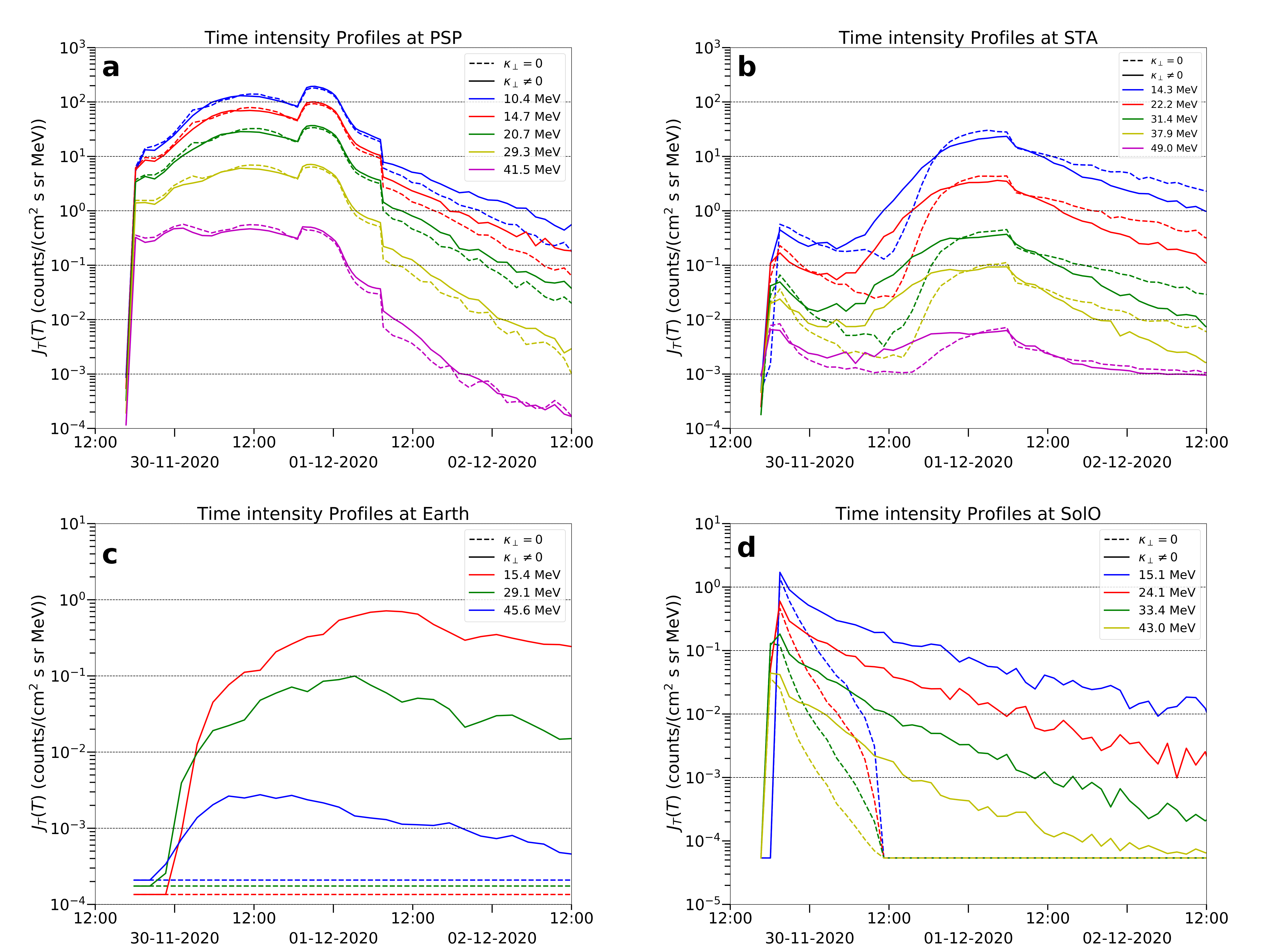}  
\caption{The comparison of time intensity profiles at PSP (a), STA (b), Earth (c) and SolO (d), with and without cross-field diffusion. The solid and dashed lines show the modelled results with and without cross-field diffusion.}
\label{fig:cross-field}
\end{figure*}

To clearly understand the role of cross-field diffusion in this widespread SEP event, we compare the cases without cross-field diffusion to the modelled results with cross-field diffusion. Figure~\ref{fig:cross-field} shows the comparison of time intensity profiles at PSP, STA, Earth and SolO. As shown in Fig.~\ref{fig:cross-field}a, since PSP is well connected to the strong shock during the event, cross-field diffusion has little effect on the time-intensity profiles. Figure~\ref{fig:cross-field}b shows time profiles at STA. There is a significant dip around 2020-11-30 12:00 UT in the case of $\kappa_{\perp} = 0$ corresponding to the weak western shock flank that barely accelerates particles. This suggests the extra proton intensity in the case of $\kappa_{\perp} \neq 0$ is contributed via cross-field diffusion. We note that the proton intensity of the decay phase in the case of $\kappa_{\perp} = 0$ is higher than the case of $\kappa_{\perp} \neq 0$, which indicates the cross-field diffusion can also reduce proton intensity when the observer already has a good magnetic connection to shock. Similar results can also be seen in \cite{Hu+etal+2018}. Figure~\ref{fig:cross-field}c plots the comparison of time profiles at Earth. As we discussed earlier, Earth is not connected to the shock in this period, so there is no enhancement of proton intensity in the case of $\kappa_{\perp} = 0$. In Fig.~\ref{fig:cross-field}d, proton intensity with $\kappa_{\perp} = 0$ at SolO shows a rapid decay to the background intensity in the first 20 hours because SolO can not connect to the shock later on. Both cases at Earth and SolO strongly suggest cross-field diffusion plays a crucial role when those observers have a poor magnetic connection to the shock.

\section{Summary and Conclusion}\label{sec:conclu}
In this paper, we present the model calculation of the 2020 November 29 SEP event by combining the EUHFORIA and the iPATH model. An important finding of this work is that the shock geometry and shock parameters are significantly regulated by the non-uniform background solar wind. This is nicely captured by the EUHFORIA code. Because the background solar wind is non-uniform, the expansion of the CME is also asymmetric, leading to deflection and deformation of the CME-driven shock. Consequently, the history of shock acceleration also becomes more asymmetric due to the expansion.  For observers with a good magnetic connection to the shock, such as PSP and STA, the history of shock acceleration plays a key role in the variation of the time intensity profiles. Besides PSP and STA, we also model proton time profiles at Earth and SolO. Both Earth and SolO are not magnetically connected to the shock in the simulation. The observed time profiles at these locations, are however, very different. At Earth, the time intensities rise gradually, showing a typical behavior for a not-well-connected observer. This agrees with the EUHFORIA simulation where Earth is not connected to the shock for the entire event.  The SEP observed at the Earth must experience cross-field transport. Under a reasonable choice of the ratio of the perpendicular diffusion coefficient to the parallel diffusion coefficient, our modelled time intensity profiles are in good agreement with observations, suggesting that cross-field diffusion is the dominant factor for the SEP event observed at the Earth. Cross-field diffusion is important to model wide spreading SEP events that are observed simultaneously by multiple observers. Although we utilize the same choice of diffusion coefficients in this work, recent studies by \citet{Ding+2020,Li2021} suggested that perpendicular diffusion can vary from event to event and can have an intra-event longitudinal dependence.  At SolO, the time profiles show prompt rises. To model SolO observations, we assume a largely distorted magnetic field line,  caused by a SIR,  to provide a better magnetic connection to the source region. In the work of \citet{Ding+2020}, they also suggested the distorted magnetic field is important in understanding the large SEP event of 2017 September 10. Comparing cases of $\kappa_{\perp} \neq 0$ and $\kappa_{\perp} = 0$ at SolO, we find that  the decay phase of time intensity profiles is sensitive to the cross-field diffusion and the observations agree better to the case with a cross-field diffusion. Therefore, we suggest that cross-field diffusion and distorted magnetic field line are both important to understand the time intensity profiles observed at SolO. We note that both effects can explain observations of SEPs in a seemingly ``no-connection" situation, and it 
can be hard to resolve the ambiguity between distorted field lines and cross-field diffusion in a given SEP event.  
 
 Our key points of the 2020 November 29 SEP event  are summarized as follows:
 
  1) The history of shock acceleration, affected by the non-uniform solar wind, plays an important role in the variation of time intensity profiles for the observers that have a good magnetic connection to the shock (i.e., PSP and STA in this event). 
  
   2) The cross-field transport dominates the time intensity profiles for the observers that are not or barely connected to the shock (i.e., Earth in this event).

  3) Transient structures (e.g., SIRs and pre-CMEs) prior to the event may largely distort the interplanetary magnetic field, leading to a change of magnetic connection. In this event, a good magnetic connection between SolO and the CME-driven shock may result from large curved magnetic field lines produced by a SIR-like disturbance. 
  
  This modelling work of the first widespread SEP event of solar cycle 25 on 2020 November 29 demonstrates the importance of extended shock acceleration and the topology of the interplanetary magnetic field in understanding the variation of SEP time intensity profiles and offers an explanation for widespread SEP events. Nevertheless, there remain some improvements that need to be implemented in the future. EUHFORIA considers a data-driven inner boundary to get a steady-state background solar wind, hence transient solar wind structures are neglected. A real-time inner boundary is necessary to be included in EUHFORIA. The iPATH model  considers the evolution of the shock wave in a data-driven background solar wind, but the model does not take into account the data-driven solar wind and magnetic field in the transport module. For simplicity, the IMF is assumed to be of Parker field in the iPATH model. However, using an arbitrary IMF resulting from a realistic solar wind can be incorporated into the iPATH model. We note that examining the transport of SEPs in the non-Parker field can be important in understanding not only energetic particle events associated with CMEs, but also events associated with SIRs. The latter has recently been investigated by \citet{wijsen2019modelling} using the PARADISE code. Finally, we point out that the inner boundary is at $0.1$ au in the EUHFORIA model, and hence particle acceleration in the low corona is neglected. Coupling to a coronal MHD model in EUHFORIA 2.0 \citep{poedts2020european} with the iPATH model, in a similar fashion as done in \cite{Li2021} for the AWSoM and iPATH, will be pursued in the future.

 Forecasting SEP events is a central topic of space weather research. This study demonstrates a promising and powerful approach to understand large SEP events by coupling the EUHFORIA and the iPATH model. As  solar activities increase intensively in solar cycle 25,  we expect to examine more SEP events, and improve the capability of the EUHFORIA and the iPATH model. 

\begin{acknowledgements}
PSP,  SolO, STEREO-A and SOHO in-situ data are publicly available at NASA’s Coordinated Data Analysis Web (CDAWeb) database (\url{https://cdaweb.sci.gsfc.nasa.gov/index.html/}). 
N.W.\ acknowledges support from the NASA program NNH17ZDA001N-LWS and from the Research Foundation -- Flanders (FWO -- Vlaanderen, fellowship no.\ 1184319N).
This research has received funding from the European Union’s Horizon 2020 research and innovation programme under grant agreement No 870405 (EUHFORIA 2.0) and the ESA project "Heliospheric modelling techniques“ (Contract No. 4000133080/20/NL/CRS).
At UAH, this work is supported in part by NASA grants 80NSSC19K0075, 80NSSC21K1814,  80NSSC20K1239, and 80NSSC22K0268.
These results were also obtained in the framework of the projects C14/19/089  (C1 project Internal Funds KU Leuven), G.0D07.19N  (FWO-Vlaanderen), SIDC Data Exploitation (ESA Prodex-12), and Belspo project B2/191/P1/SWiM.
For the computations we used the infrastructure of the VSC–Flemish Supercomputer Center, funded by the Hercules foundation and the Flemish Government–department EWI.

\end{acknowledgements}

%
  \bibliographystyle{aa} 
  \bibliography{joint_refs}
%

\end{document}